# Band Gap Opening in Stanene Induced by Patterned B-N Doping

Priyanka Garg,[†] Indrani Choudhuri,[†] Arup Mahata,[†] Biswarup Pathak,[†,#,*]

[†]Discipline of Chemistry and [#]Discipline of Metallurgy Engineering and Materials Science, Indian Institute of Technology (IIT) Indore, Indore. M.P. 453552, India
Email: biswarup@iiti.ac.in

**Abstract:**

Stanene is a quantum spin hall insulator and a promising material for electronic and optoelectronic devices. Density functional theory (DFT) calculations are performed to study the band gap opening in stanene by elemental mono- (B, N) and co-doping (B-N). Different patterned B-N co-doping is studied to change the electronic properties in stanene. A patterned B-N co-doping opens the band gap in stanene and the semiconducting nature persists with strain. Molecular dynamics (MD) simulations are performed to confirm the thermal stability of such doped system. The stress-strain study indicates that such doped system is as stable as pure stanene. Our work function calculations show that stanene and doped stanene has lower work function than graphene and thus promising material for photocatalysis and electronic devices.

## 1. Introduction

Two-dimensional (2D) nanostructures have attracted considerable attention because of their exceptional electronic and optical properties.[1-7] After, the synthesis of graphene, the heavy group-IV elements (Si, Ge, and Sn) are receiving considerable attention due to their similar electronic properties. For example, linear band dispersion at the Fermi energy, which leads to charge carriers, behaves like massless Dirac Fermions.[3] These massless Dirac Fermions propagate with a very good Fermi velocity and show high carrier mobilities at room temperature.[5]

So it is very vital to synthesize atomically thin group-IV materials. Over the past few years, group-IV monolayers have been experimentally realized using different experimental techniques.[8-11] Two-dimensional buckled silicene and germanene have been synthesized using molecular beam epitaxy method.[8,12] Similarly a monolayer of Pb has been deposited on a metal surface using ultrahigh vacuum techniques.[13-14] In 2011, Garcia et al. have theoretically predicted various allotropes of atomically thin Sn and demonstrated similar electronic properties like graphene.[15] Later, Zhang and co-workers proposed a monolayer of tin as stanene in analogy with graphene and silicene.[16] It took four more years to realize experimentally as stanene has been synthesized on the $Bi_2Te_3(111)$ surface via an epitaxial growth.[17] Furthermore, Gao et al. have studied the epitaxial growth of stanene over the Ag (111) surface and their study revealed that the epitaxial growth of stanene is preferred due to a low energy barrier. More importantly, the hexagonal structure preserves its symmetry conserved even after the epitaxial stanene monolayer is removed from the Ag (111) surface.[18] So, stanene is the newest addition to the atomically thin group-IV family.

Stanene is a bi-atomic layer of α-Sn(111), which has a graphene-like honeycomb structure.[18] However, the structure is most stable with a buckled configuration due to weak π-π bonding between Sn atoms.[16] Theoretical studies demonstrated that stanene has a zero band gap without inclusion of spin orbit coupling (SOC), whereas it shows a band gap of 0.1 eV after inclusion of SOC.[19] This is an interesting finding within the group-IV family with a band gap of 0.1 eV as the next member (Pb) is metallic. Stanene behaves like a quantum spin hall insulator (QSHI) and its derivatives show properties of large-gap quantum spin Hall insulators (QSHIs).[16] Thus, such QSHIs show dissipationless conductor at room temperature.[16] Furthermore, such QSHIs show excellent thermoelectric.[20] and quantum anomalous Hall (QAH) effect near room temperature.[21] On the other hand, doped QSHI materials show time-reversal-invariant topological superconductivity.[22]

Thus, doped stanene could be a promising material for various applications. On the other hand, the zero-band gap in stanene also limits its applications towards semiconductor based microelectronic devices.[23] Therefore, the expectation of a band gap opening in stanene has spurred intense scientific interests. Various approaches have been taken to modulate the electronic properties of stanene.[24-28] Tang et al.[27] demonstrated that functionalization on stanene can open up the gap. Furthermore, the substrate plays an important role to tune the electronic properties of stanene due to the strong interaction.[25,29] Electronic properties of stanene are affected by the substrate because it induces band inversion in the bonding and antibonding states of stanene, which opens up the band gap in stanene.[27] Furthermore, van der Waals heterostructure of stanene modulates its electronic properties.[30] Recently, Wang et al. demonstrated a heterostructure of stanene with h-BN sheet induces a tunable gap in stanene.[30] On the other hand, doping is also an efficient way to tune the electronic properties of 2D nanosheets by breaking the symmetry of the sublattice.[31-34]

In this work, we have performed a systematic study to tune the electronic properties of stanene. Various mono- (B, N) and co-doped (B-N) structures are considered to modulate the electronic properties of stanene as such mono and co-doping found to be very effective to tune the electronic properties of group-IV family.[32,35-40]

In the past few years, mono (B/N) and co-doped (B-N) graphene systems are experimentally synthesized. Moreover, they could achieve doping concentration as high as 13% for B doping graphene, 15.60% for N doped graphene, and 27% for co-doping.[41-43] Chang et al.[42] synthesized mono- and co-doping in graphene using chemical vapour deposition technique.[44-45] Thus, we believe that B, N, and B-N co-doped stanene can be experimentally synthesized using similar techniques.

## 2. Computational Details

We have used the Vienna Ab Initio Simulation Package (VASP)[46] to perform all the calculations. The exchange-correlation potential is treated in the level of the GGA using Perdew-Burke-Ernzerhof (GGA-PBE).[47] For describing the electronic wave function, the projected augmented wave (PAW)[48] method is employed using an energy cut-off of 500 eV. Self-consistency is achieved with convergence tolerance set to $10^{-4}$ eV and $10^{-2}$ eV/Å for total energy and force calculations, respectively. The first Brillouin zone of the supercell (5×5) structure is sampled with a 5×5×1 Gamma-pack k-point grid for geometry optimization and 15×15×1 for the density of state calculation. About 20 Å of vacuum is employed in the z-direction to avoid any interaction between the periodic layers. Furthermore, a hybrid functional method (HSE06)[49] is used to verify the electronic properties of the doped systems. Molecular dynamics simulation (MD Simulation) is controlled by the Nosé thermostat model,[50] which carried out at 300 K temperature for 20 picoseconds (ps) with a time step of 1 femtosecond (fs). Bader charge analysis[51-52] is done using the Henkelman code with near-grid algorithm refine-edge method.[53-54] Phonon properties are calculated using density functional perturbation theory (DFPT)[55] as implemented in the VASP and phonon dispersion calculations are carried out using Phonopy code.[56]

## 3. Results and Discussion

## 3.1. Stanene

A two-atom hexagonal unit cell is considered for the stanene structure. The computed lattice parameters in stanene are a = b = 4.67 Å and the Sn–Sn bond length is 2.83 Å, which are consistent with the previous reported value of 2.82 Å.[57] Our optimized structure shows that stanene has a buckling of 0.87 Å, which agrees with the previous reported value of 0.85 Å[16,17,19] (Figure 1). Earlier report shows that the buckled form of stanene is the most stable, which contrasts with planar geometry of graphene and the buckling is due to the weak π-π interaction in stanene.[16]

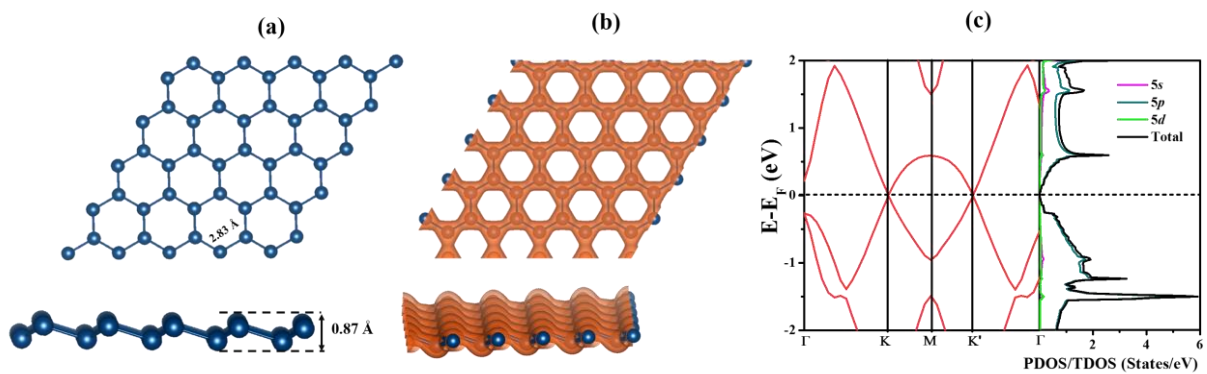

**Figure 1**(a) Optimized structure of stanene (front and side views and buckling height of 0.87 Å), (b) total electron density surface (front and side views; Isosurface value: 0.03 e.Å$^{-3}$), and (c) band structure and total density of states of stanene. Here, the Fermi level is set to zero and indicated by a black dashed line.

The Bader charge analysis shows that each Sn atom possesses 3.99 |e| amount of charge, which shows the covalent nature of bonding.[17,58] Further, the electron density plot (Figures 1b) indicates that charges are highly delocalized over the surface. On the other hand, the electronic structure of pure stanene is another interesting finding in the field of 2D materials. Thus, the total and partial density of states (TDOS/PDOS) of stanene (Figure 1c) are plotted to understand its electronic properties. The valance (VB) and conduction bands (CB) of stanene consist of Sn *5p* orbitals.[57] Further, figure 1c shows that stanene is a gapless material with the Dirac cones at K points. However, the band gap opens as we include the spin orbit coupling (SOC) in our electronic structural calculations. Stanene shows a band gap of 0.1 eV

(Figure S1(a), Supporting Information) after inclusion of SOC, which is very much in agreement with the previous reports on stanene.[24, 59]

### 3.2. Doping in Stanene

Substitutional elemental doping is an effective way to tune the electronic properties of materials. Being the fourth member of carbon family, stanene has similar electronic properties like graphene. So, the band gap opening in stanene could be possible by elemental mono- (B, N) and co-doping (B-N). The two-atom unit cell is extended to a 5×5 supercell of 50 atoms to achieve lower doping concentrations. Sn is a group IV element, and if we dope a group III/V (B/N) element, then it induces a hole/electron in the system.[35] Thus, it will break the symmetry of the sub lattice, which in turn, might change the electronic properties of the sheet.

### 3.2.1. B@Stanene

Boron (B) doped systems are widely investigated and found to be promising for electronic and related applications.[60] Here, we have considered 2% of B doping i.e. one B atom is doped in the 50 atomic unit cell of stanene. The previous report demonstrated that 2% B doping is favourable in graphene[45,61] due to the lower concentration that maintains the symmetry of sub lattice. However, higher doping concentration modifies the structure of graphene.[60] So, here we have considered only lower doping concentration for our study in stanene. The B@stanene structure buckles like stanene though the buckling in B@stanene is higher than in stanene (Table 1). Furthermore, the structure distorts more at the B-centre due to the difference in atomic radius. Thereby, the Sn-B distance is 2.24 Å compared to Sn-Sn bond distance of 2.83 Å (Figure 2a). Our charge density analysis (Figure 2b) shows that electron density is maximum at B-site and this can be attributed to the electron deficient nature of B atom. Furthermore, this can be confirmed from the Bader charge analysis, which shows that the B is negatively charged (-0.53 $|e|$), while the neighbouring Sn are positively charged (+0.18, +0.18 and 0.17 |e|). The TDOS/PDOS of B@stanene are plotted (Figure 2c) to understand the change in electronic properties due to B-doping. The electronic structure (Figure 2c) shows that the Fermi level is shifted towards the valance band edge compared to in pure stanene and this confirms that the B@stanene is p-type of material. Further, such doping shifts the Dirac cone by 0.26 eV (Figure 2c) from the Fermi level. Figure 2c shows that the boron $2p$ orbitals appear at the Fermi level along with Sn $5p$ orbitals. It indicates that B@stanene is a metallic system.

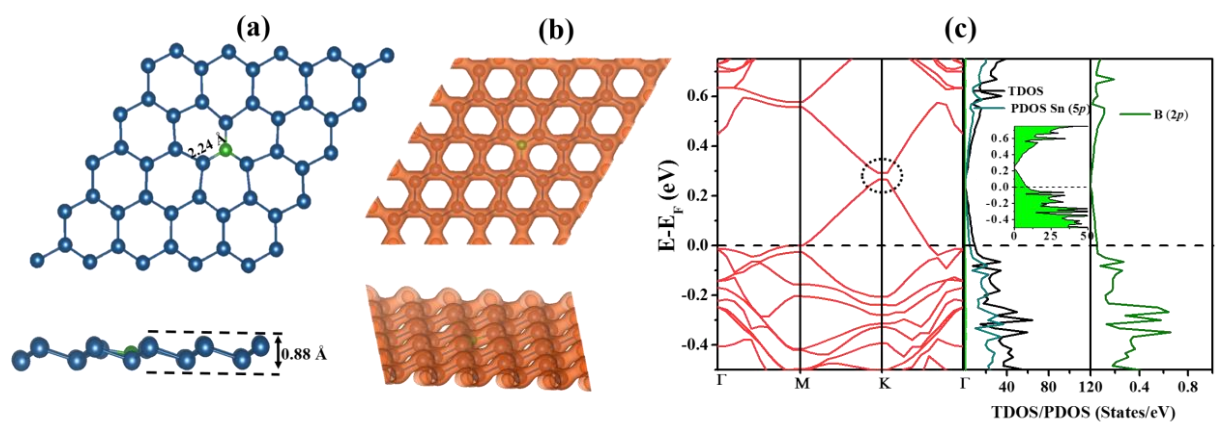

**Figure 2:** (a) Optimizes structure of B@stanene (front and side views; buckling height 0.88 Å), (b) total electron density surface (front and side views; Isosurface: 0.03 e. Å$^{-3}$), and (c) band structure and total/partial density of states of B@stanene. Here, green and cyan balls represent B and Sn atoms, respectively. The Fermi level is set to zero and indicated by a black dashed line.

However, there is a band gap of 0.02 eV above the Fermi level, which indicates that the B@stanene could be a degenerate semiconductor. Furthermore, we have calculated electronic properties of B@stanene using HSE06[49] level of theory to verify the degenerate semiconducting nature of B@stanene. The HSE06 level of theory (Figure S3, Supporting Information) confirms the degenerate semiconducting nature of B@stanene. However, the band gap above the Fermi level is higher (0.28 eV) than achieved by GGA (0.02 eV) level of theory and this is understandable as GGA underestimates the band gap.[47] The degenerate semiconducting nature of B@stanene persists (Figure S1(b), Supporting Information) even after the effect of SOC. However, the Dirac cone disappears after the inclusion of SOC effect. Thus, the electronic properties of B@stanene change due to the SOC effect. In fact, the electronic properties of such system can be tuned by tuning the Dirac point using the gate voltage.[62] Furthermore, such type of features are observed in B@graphene too.[35,63] Thus, we predict B@stanene could be a promising material for electronic and optoelectronic devices.

**3.2.2. N@stanene**

Nitrogen doped graphene is a promising material for energy, catalysis and electronic devices.[64] Thus, we predict that N@stanene could show similar electronic properties like graphene. Here, we have doped one N atom in the 50 atomic unitcell of Sn, which corresponds to 2% of N-doping concentration that is favorable in N@graphene.[41,43]

We find that the buckling is maximum (0.98 Å) in N@stanene compared to in B@stanene and stanene. Furthermore, the structure shows more distortion around the N-center and this can be understood from the Sn-N bond distance (2.14 Å) in comparison to Sn-Sn bond distance (2.83 Å) in stanene (Figure 3a). The total electron density surface (Figure 3b) indicates the covalent nature in stanene and the bonding is disturbed due to N doping. The Bader charge analysis shows N atom gains -1.27 |$e$| more negative charge from surrounding Sn atoms and thus Sn atoms are positively charged 0.48-0.49 |$e$|.

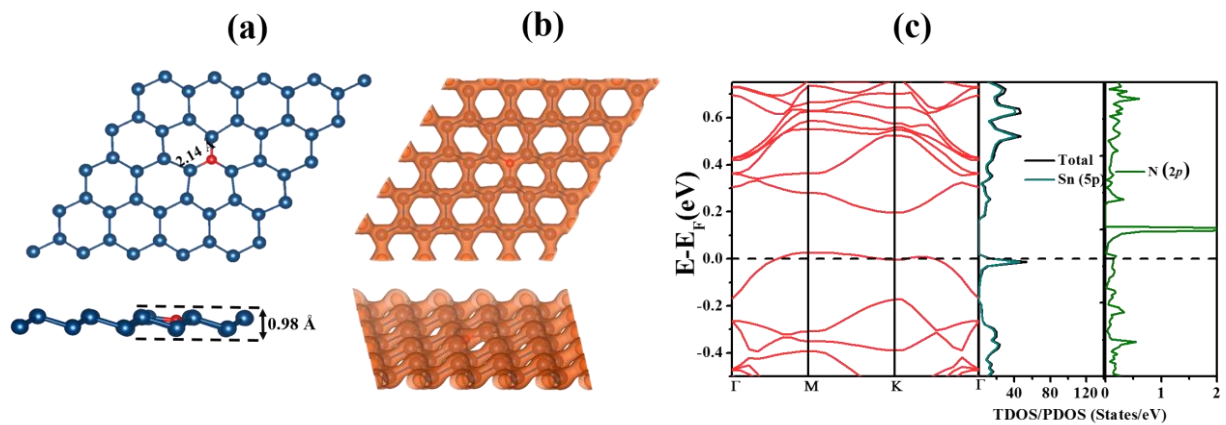

**Figure 3:** (a) Structure (front and side views; buckling height 0.98 Å), (b) total electron density surface (front and side views; Isosurface: 0.03 e. Å$^{-3}$), (c) band structure and total/partial density of states of N@stanene. Here, cyan and red balls represent Sn and N, respectively and the Fermi level is set to zero and indicated by the black dashed line.

To understand the electronic properties of N@stanene, the TDOS/PDOS of N@stanene are plotted (Figure 3c) and compared with pure and B@stanene. The electronic structure (Figure 3c) shows that the Fermi level is shifted towards the conduction band edge compared to pure stanene and this confirms that the N@stanene is an n-type of system. Furthermore, N@stanene is different from B@stanene, as the Dirac cone exists in B@stanene, whereas vanishes in N@stanene. Thus, the N@stanene destroys the linear dispersion of π bands (Dirac cones) at the Fermi level (Figure 3c). This is very much consistent with previous report on N doped silicene.[36] Figure 3c shows that the nitrogen 2$p$ orbitals appear at the Fermi level along with Sn 5$p$ orbitals. Thus, N@stanene could be a metallic. However, there is a band gap of 0.17 eV above the Fermi energy, which indicates that the N@stanene is a degenerate semiconductor and the electronic properties of such system, can be tuned by tuning the gate voltage.[61] Similar to B@stanene, the degenerate semiconductor nature of N@stanene is further confirmed by HSE06 level of theory. Though the observed band gap above the Fermi level is higher (0.21 eV) (Figure S3, Supporting Information) than that obtained by GGA (0.17 eV) level of theory. Similar properties are observed in B/N@graphene too.[35] Thus, we predict N@stanene could be a promising material like N@graphene for electronic and other applications. In case of N@stanene, the total density of states is similar at the Fermi level even after the inclusion of SOC effect (Figure S1(c), Supporting Information).

### 3.2.3. B-N@stanene

As mono-doping found not to be beneficial for band gap opening in stanene, we have considered B-N co-doping for changing the electronic properties in stanene. Furthermore, such B-N co-doping has been experimentally achieved in graphene and reported to be promising for various applications.[43-45] More importantly, B-N co-doping induces both the electron and hole simultaneously.[35] So, there is a possibility to open the gap in stanene. For this, we have considered different doping patterns and have been studied in detail. In case of B-N co-doping, one B and one N atom is doped in the 50-atom supercell of stanene, which corresponds to 4% of B-N doping.

**Doping Patterns:**

We have adopted two distinctly different possibilities for B and N co-doping in stanene, where B and N atoms are either doped in a closest or furthest possible ways. Keeping these in mind, we have considered four different doping patterns (pattern-a, pattern-b, pattern-c, and pattern-d) for B-N co-doping. In case of pattern-a (Figure 4a), we have co-doped B and N as a BN-domain, where B and N atoms are sitting next to each other. However, in the optimized structure, B and N atoms are not covalently bonded. This can be further confirmed from the total electron density surface analysis (Figure 4a). It shows that the electron densities are not delocalized along the B-N bond. This suggests that the B-N domain might not be favoured in stanene. Furthermore, the Bader charge analysis on B-N@stanene shows that the more charge is localized on N than on B. However, both the atoms (B and N) are negatively charged in B-N@stanene. Thus, it indicates that both B and N withdraw electron from Sn in stanene. Thus, both (B and N in pattern-a) the negatively charged atoms repel each other. To understand this, we have compared Bader charges of B and N atoms in B-N@graphene, where B and N atoms are doped in the similar manner as in pattern-a. For comparison, we have calculated Bader charges in B-N monolayer too. In contrast, B atom is positively charged in B-N@graphene (+0.30 |$e$|) and B-N monolayer (+2.20 |$e$|), whereas N atom holds negative (-2.20 and -1.47 |$e$|) charge in them. However, in B-N@stanene, the charges on B and N atoms are -0.38 and -1.15 |$e$|, respectively. This occurs due to the relative electronegativity (N > C > B > Sn) of these atoms. Therefore, due to the presence of negative charge on N and B atoms, they repel and do not form a B-N bond.

For pattern-b (Figure 4b), B and N atoms are doped in the same hexagon of stanene but sitting just opposite to each other. In case of pattern-c (Figure 4c), B and N atoms are doped in an alternate manner and in that case one Sn atom is present between B and N atoms (Figure 4c). For pattern-d (Figure 4d), B and N atoms are doped in such a way that the distance is maximum in the 50-atomic unit of supercell. However, the pattern-b structure (Figure 4b) shows lesser distortion and less buckling compared to the pattern-a structure. The total electron density surface plot of pattern-b structure indicates that B and N are strongly bonded with neighbouring Sn atoms. In the pattern-c (Figure 4c), B and N atoms are doped in an alternate manner and thus one Sn atom is present between B and N atoms (Figure 4c). Our total electron density surface plot shows that it has similar bonding pattern as in the pattern-b structure (Figure 4b-c). Another pattern (pattern-d, Figure 4d) is considered, where B and N atoms are doped in such manner that the distance between B and N is maximum in the 50-atomic supercell. The optimized structure of pattern-d shows that, it has lowest buckling among all other B-N doped structures (Table 1). The lower buckling of pattern-d could be due to the maximum distance between B and N. Total electron density plot (Figure 4d) indicates that the charges are localized on B and N atoms, which is very similar to other B-N@stanene structures.

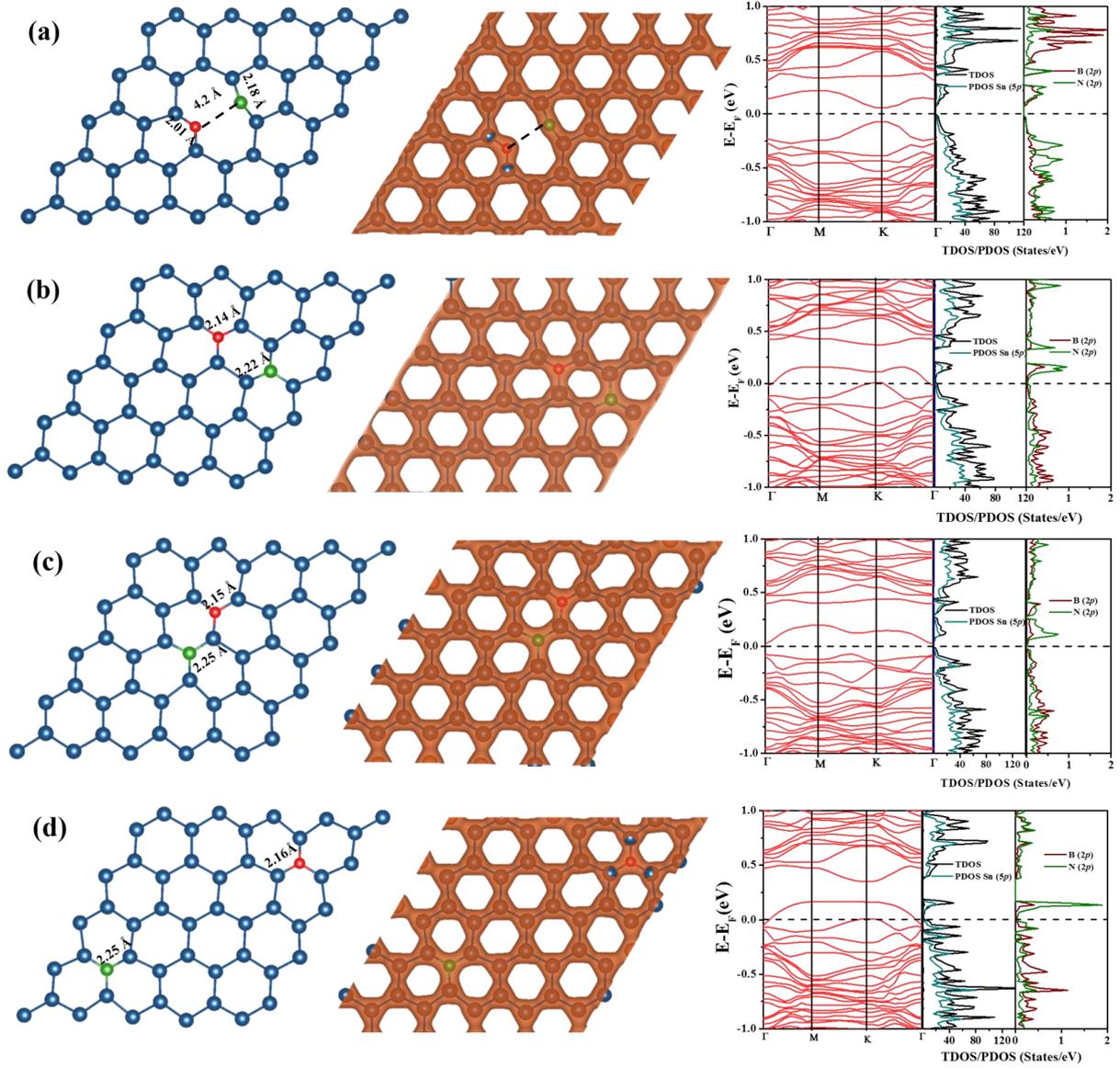

**Figure 4:** Structure, total electron density (Isosurface value is 0.03 e.Å$^{-3}$) and electronic structure (band structure and TDOS/PDOS) of (a) pattern-a, (b) pattern-b, (c) pattern-c and (d) pattern-d B-N@stanene. The Fermi level is shifted to zero and indicated by a black dashed line.

Further, we have calculated electronic properties (Figure 4) of different patterned B-N@stanene. Figure 4a shows that there is a band gap opening (0.08 eV) at the Fermi energy, which corresponds to pattern-a type of B-N doped system. However, other B-N@stanene

structures do not open the gap at the Fermi level (Figure 4b-d). In pattern-a, the distance between B and N is 4.2 Å, which is very high compared to B-N bond distance (1.45 Å) in boron nitride. Furthermore, our band structure plot (Figure 4a) shows that it has a direct band gap of 0.08 eV along the M → K path of the Brillouin zone. The electronic structure calculated using HSE06 level of theory shows a band gap of 0.22 eV (Figure S3(c), Supporting Information) in the pattern-a structure. Thus, the HSE06 level of theory confirms that a patterned B-N doping induces a semiconducting property in stanene. The electronic properties of differently patterned B-N doped stanene systems are studied in the presence of SOC effect. We find that the electronic properties of pattern-a type B-N doped system does not affected due to the SOC effect as it maintains its semiconducting nature (Figure 4a). However, other B-N doped systems (pattern-b-d) show band gap opening in the presence of SOC effect [Figure S2 (a-d), Supporting Information]. In fact, the band gap opening is significant (0.10-0.13 eV) due to SOC effect as observed in stanene. Thus, these systems may behave like a quantum spin hall insulator[16] and may shed some light in the field of topological

. Thus, we plan to do a detailed investigation on these B-N doped systems and on similar systems to understand whether such materials can behave like quantum spin hall insulators.

### 3.3. Formation Energy

The formation energy (per supercell) calculations (see Text S1, Supporting Information) are performed for B@, N@ and B-N@stanene to investigate the thermodynamic stability of such doped systems.[35,63] Based on our formation energy calculations, we find that N@stanene formation is favorable over B@stanene and B-N@stanene (Table 1).[65] The size of N (0.75 Å) is smaller than B (0.82 Å) and more electronegative nature of nitrogen atom which added an ionic binding in initial co-valent binding.[65] The lower value of formation energy by N doping is also achieved in silicene which is comparable with our results in stanene.[65] Among all the B-N@stanene, the formation of pattern-b is energetically favourable over other B-N doped structures (Table 1). However, such B-N co-doped systems with high formation energy value can be synthesized. For example, the calculated formation energy of 15.60% B doping (7.56 eV) in graphene is much higher than N (2.68 eV) doping in graphene.[35,62] However, 13% B doping graphene has been synthesized.[66-67,68] On the other hand, the calculated formation energy (per supercell) for 31% B-N co-doped graphene is 4.08 eV, which is significantly lower than the formation energy of 15.60% B-doped graphene system.[35] However, 27% B-N co-doped graphene[43] has been synthesized. Thus, we predict that B-N co-doped stanene can be experimentally realizable using similar experimental techniques.

**Table 1**: Doping percentage (%), formation energy/supercell ($E_f$), buckling height (h), important bond lengths (Sn-B and Sn-N), and band gap ($E_g$) of pure, mono, and co-doped stanene are given.

| Doped System | % | $E_f$ (eV) | h (Å) | $d_{Sn-B}$ (Å) | $d_{Sn-N}$ (Å) | $E_g$ (eV) |
|---|---|---|---|---|---|---|
| **Pure-stanene** | - | - | 0.87 | 2.83(Sn-Sn) | - | 0 |
| **B@stanene** | 2 | 1.84 | 0.88 | 2.24 | - | - |
| **N@stanene** | 2 | 1.32 | 0.98 | - | 2.14 | - |
| **B-N@stanene** | | | | | | |
| Pattern-a | 4 | 4.40 | 1.09 | 2.18 | 2.01 | 0.08 |
| Pattern-b | 4 | 2.79 | 1.01 | 2.22 | 2.14 | - |
| Pattern-c | 4 | 3.15 | 1.09 | 2.25 | 2.15 | - |
| Pattern-d | 4 | 3.15 | 0.99 | 2.25 | 2.16 | - |

### 3.4. Thermal Stability

Thermal stability of B@, N@ and B-N@stanene is evaluated by performing ab initio molecular dynamics simulation (AIMD) using Nosé thermostat as implemented in VASP.[46] Simulations are carried out on the 5×5 supercell of the doped stanene, using an NVT ensemble at 300 K with a time step of 1 femto second (fs) for 20 pico second (ps). The energy fluctuation is not very high throughout the simulation. Thus, we can say that all the doped systems are stable at room temperature (Figure S4-S5, Supporting Information).

### 3.5. Dynamic Stability & Properties.

Based on our electron structure calculations, we find that a patterned (pattern-a) B-N doping in stanene is important for band gap opening. So, we have calculated the dynamic stability of doped stanene structure and compared with pure stanene using phonon dispersion calculations (Figure 5a-b). These calculations are carried out using density functional perturbation theory[54] as implemented in the VASP package.[46] The Phonopy code[56] is used to plot the different modes of phonon. The absence of imaginary modes in the entire Brillouin zone confirms that stanene is a dynamically stable structure.[69] However, we have found a small imaginary frequency of ~10$i$ for B@ (10.6$i$, Figure S6a), N@ (8.8$i$, Figure S6b), Supporting Information) and 4%B-N@ stanene pattern-a (9.7$i$, Figure 5b) structure. Such a small imaginary value can be neglected as reported earlier.[70-71] Furthermore, we have calculated phonon dispersion for other B-N doped structures (pattern-b, pattern-c and pattern-d) and their respective phonon dispersion plots are shown in Figure S7 (b-d) of Supporting Information. The phonon dispersion plots of all the B-N doped systems indicate that the pattern-a is the most stable structure among all the B-N doped structures. The calculated imaginary frequencies of pattern-b, pattern-c and pattern-d systems are 29.5$i$ (Γ→M), 21$i$ (at K) and 103.7$i$ (at Γ), respectively. In the case of pure stanene, there are three distinct acoustic and optical phonon branches (Figure 5a) in the phonon spectra, which is similar to graphene.[69] Furthermore, stanene is reported to be a good thermoelectric material and has lower thermal conductivity than other 2D materials like graphene.[72] This is also due to the large gap in phonon density of states between acoustic and optical modes. Thus, stanene is a good material for thermoelectric device[69] and shows better thermoelectric properties than graphene. In the past few years, many studies illustrated that due to doping and some other defects can reduce the thermal conductivity of material.[73-74] Thus, we assume that such B-N co-doping in stanene might lower the thermal conductivity in stanene. Moreover, such B-N@stanene can be useful for electronic and optoelectronic devices.[75]

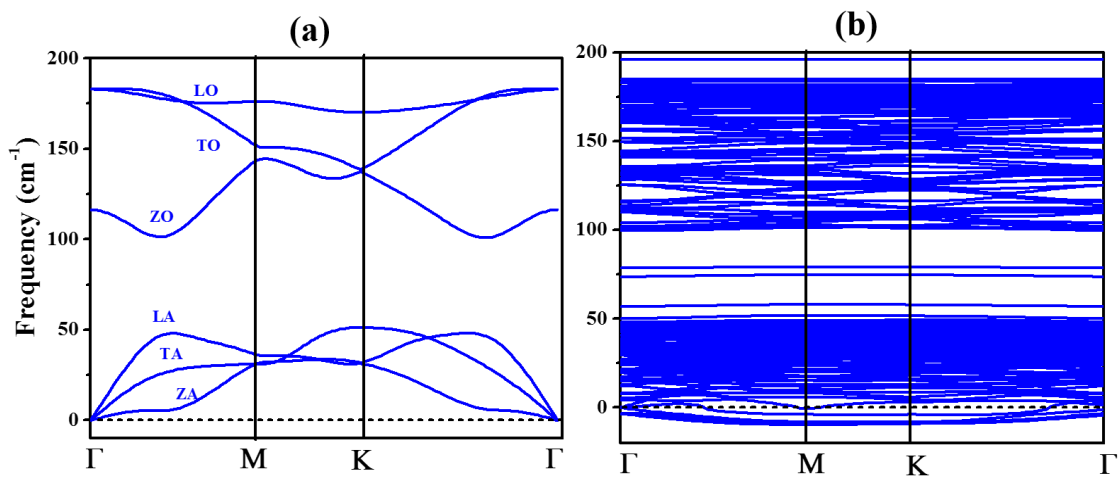

**Figure 5:** (a) Phonon dispersion of a two-atom unitcell of stanene, and (b) 5×5 supercell (50 atom) of B-N@stanene (pattern-a). Here Z, T and L represent the z-axis, transverse and longitudinal, respectively. On the other hand, A and O represent acoustic and optical phonon modes, respectively.

### 3.6. Mechanical Stability

As we have found that the B-N co-doping is efficient for the band gap opening in stanene. Thereby, we have performed mechanical stability of such system for their practical usages. Strain technology is very important for tuning the structural and electronic properties of materials. Hence, structural stability is very important under strain. To evaluate its mechanical stability and change in electronic properties under strain, we have applied in-plane uniaxial and biaxial tensile strains on the system. The following formula is used for calculating the % of strain applied.[76-78]

$$\%\text{Strain} = \frac{a-a_1}{a} \times 100 \qquad (1)$$

where a and $a_1$ are the lattice constants of the monolayer sheet before and after the strain.

Atomic positions are relaxed at each tensile strain until the forces on each atom are less than 0.01 eV/Å. After each applied strain, the value of stress is the response of system against strain. All the values of stress can be rescaled by $Z/d^0$ to get the approximate stress, where Z is the vacuum length in z-direction and $d^0$ is the thickness (3.20 Å) of the system[77,79] Elastic limit is calculated from the stress-strain relationship (Figure 6a) under tensile strain.

The effect of in-plane uniaxial and biaxial tensile strains on the B-N@stanene (pattern-a) is studied to evaluate its mechanical properties. The maximum stress occurs at 10% strain for uniaxial as well as biaxial strains. The elastic limit of the system is at 0.10 strain with a maximum stress of 4.18 GPa. These data are very much comparable with the previous report on pure stanene.[80] Thus, it indicates that the mechanically stability of B-N@stanene is comparable with pure stanene.

Furthermore, the electronic properties of B-N@stanene are evaluated under uniaxial/bi-axial tensile/compressive ($\pm 1$ to $\pm 5$) strains (Figure 6b-e, Figure S8-S11, Supporting Information). Their respective results are given in Table S1 (Supporting Information). During compression (1-5%), the average Sn-B bond length reduces from 2.18 to 2.16 Å. Similarly, under tensile strain (1-5%), the Sn-B bond length elongates from 2.18 to 2.20 Å. However, Sn-N bond distance does not change much under strain (Table S1, Supporting Information) In all these cases, and the structure buckles significantly under compressive strain. However, the buckling is minimum under tensile strain. The variation in band gap (Table S1) under different uniaxial/biaxial strains, are represented in figure 6b. The semiconducting nature of 4% B-N@stanene in pattern-a is maintained under 4% uniaxial tensile strain and upto 5% uniaxial compressive strain. It shows that the band gap sharply increases to 0.13 eV under uniaxial compressive 1-2%. This could be due to the high buckling compared to the pure system (Figure 6b).[81]

Similarly, we have studied the effect of biaxial strain on B-N@stanene as shown in figure 6a-b and Table S1 (Supporting Information). The maximum stress occurs at 10% tensile strain and that is 2.18 GPa (Figure 6a). We have observed similar trend on stress under bi-axial strain as observed in uniaxial strain. The TDOS/PDOS of B-N@stanene at different biaxial strains illustrates that the band gap is maintained till 2% tensile strain, while maintains till 3% under compressive strain.

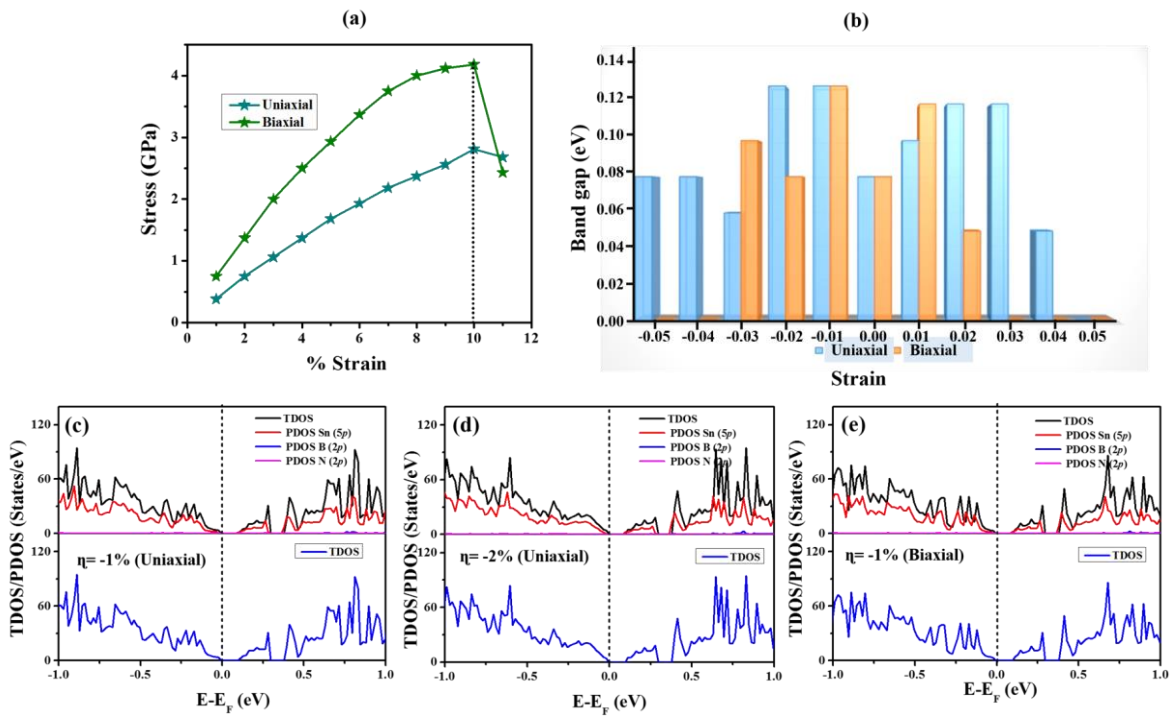

**Figure 6:** (a) Stress vs. strain relationship of 4% B-N@stanene (pattern-a). Here the black dash line indicates the maximum value of stress for in-plane uniaxial and biaxial strains. (b) Graphical representation of variation of band gap of 4% B-N@stanene (pattern-a) under uniaxial and biaxial strains. TDOS/PDOS of 4% B-N@stanene under (c) 1 % uniaxial, (d) 2% uniaxial and (e) 1% biaxial compressive strains.

The maximum value of band gap (0.13 eV) observed in biaxial 1% compressive strain (Figure 6e), whereas the system behaves like a metallic under higher compressive strain (Figure S11, Supporting Information). Thus, we find that 4% B-N@stanene (pattern-a) can retain its semiconducting nature under extreme strain (upto 5% uniaxial and biaxial strain) (Table S1, Supporting Information), which could be due to the buckling.[81]

### 3.7. Photocatalytic Properties

We have calculated the work function of the pure stanene and 4% B-N@stanene (pattern-a) to locate the Fermi position with respect to the water oxidation and reduction potentials. Work-function ($\phi$) is calculated using the following equation:[77]

$$(\phi) = E(vacuum) - E_F \qquad (2)$$

where, E (vacuum) and $E_F$ are the energy of the vacuum and the Fermi level, respectively.

In the past few years, graphene is used as a cocatalyst with wide band gap semiconductor material for improving the photocatalytic efficiency. Graphene sheet acts as an electron transfer channel, which suppress the recombination of electron and hole.[82-83] Graphene and graphene based materials serve as good photocatalytic materials owing to their high electron conductivity, mobility, adsorption capacity and specific surface area.[84] Stanene is a quantum spin hall insulator, which conducts electricity through its edges and can be 100% electrical efficient.[16] So it is expected that it could conduct dissipationless current with very low power consumption.[16] We think that stanene can be used as a co-catalyst like graphene to improve the photocatalytic properties of the photocatalyst. For this, we have calculated work functions of pure stanene, 4% B-N@stanene (pattern-a), and graphene and plotted with respect to water oxidation/reduction potential to evaluate their relative band edge potentials (Figure 7). In case of a semiconductor based photocatalyst, the position of the conduction band minimum (CBM) should be above (more negative than) the $H^+/H_2$ reduction potential (0.0 eV) and the valance band maximum (VBM) should be below (more positive than) the $O_2/H_2O$ oxidation potential (1.23 eV).[85] The calculated work function values of graphene, stanene and B-N doped stanene (pattern-a) are 4.50, 4.22 and 4.25 eV, respectively. In B-N@stanene, the valance band edge position is close to the Fermi level (-4.26 eV), whereas the conduction band edge is just above the Fermi level (-4.18 eV). This is similar to pure graphene. Our calculated work function of graphene (4.50 eV) is very much agreement with the previous experimental and theoretical reports.[86] Besides, the calculated work functions (4.22 eV) of stanene is very much in agreement with the previous report[87] and the calculated work function of 4% B-N@stanene (pattern-a) is 4.25 eV (Figure S12, Supporting Information). This shows that the work function value is lower in stanene and doped stanene compared to graphene. Stanene has a better absorption property than graphene as it absorbs light in the range of 200-300 nm compared to graphene which absorbs in the range of short-wavelength.[87] The single-layer hexagonal graphene material has been reported to improve the photocatalytic activity, where graphene acts as a reservoir for photogenerated-electrons.[88] Similar to graphene, stanene and doped stanene can act as a reservoir of photogenerated electrons as they show similar electronic properties like graphene. Moreover, it shows a linear band dispersion at the Fermi energy and thus charge carriers can behave like a massless Dirac Fermions.[3] Propagation of these massless Dirac Fermions occurred with very high Fermi velocity and display high carrier mobilities at room temperature.[5,16] Thus, stanene has a great potential for photo-related applications than graphene.[87] Moreover, achieving ultra-low work function in graphene is very important for electronics and electron emission devices.[89-91] As stanene and doped stanene have lower work function than graphene, they can be promising for electronic device technologies.[92] Thus, stanene can be used for photocatalytic application as an electron transfer channel for Z-scheme photocatalysis.[93]

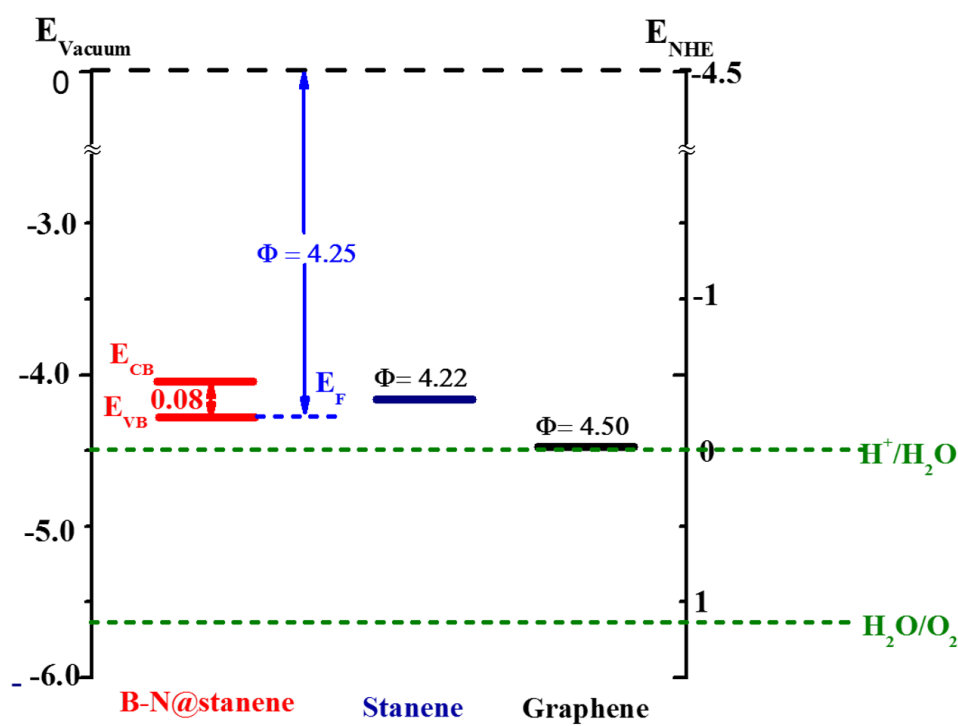

**Figure 7:** Schematic representation of band edge alignment of 4% B-N@stanene (pattern-a), stanene and graphene with respect to water oxidation and reduction potentials. Here, $E_{CB}$, $E_{VB}$, $E_F$ and $\Phi$ represent conduction band (eV), valance band (eV), Fermi energy (eV) and work-function (eV), respectively.

## 4. Conclusion

Here we have performed a systematic study on electronic, thermal, dynamical and mechanical properties of pure, mono- (B, N) and co-doped (B-N@) stanene using density functional theory (DFT) calculations. Our electronic structure calculations show that all the mono-doped structures are metallic. Thus, the *p/n*-type of doping does not open the band gap in stanene. Nevertheless, they open up the band gap above the Fermi level and thus can be used as degenerate semiconductors. However, a patterned B-N co-doping (pattern-a) opens up the band gap (0.08 eV and 0.22 eV by GGA and HSE06, respectively) in stanene and B-N@stanene behaves like a low band gap semiconductor. Interestingly, such semiconducting behaviour persists under strains. Furthermore, we find that stanene and B-N@stanene has lower work function than graphene. As achieving ultra-low work function in two dimensional materials, is very important for electronics and electron emission devices. Moreover, such B-N@stanene is energetically, dynamically, thermally and mechanically stable. So, we believe that such patterned co-doping is very important and effective for tuning the electronic and optoelectronic properties of stanene and B-N@stanene can be a potential material for thermoelectric and photo-related applications.

## 5. Acknowledgments

We thank IIT Indore for the lab and computing facilities. This work is supported by DST-SERB (EMR/2015/002057). P.G. I.C and. A.M thank MHRD for their research fellowship.

## 6. Associated Contents:

***Supporting Information:**

TDOS (total density of states) of B@stanene, N@stanene, and 4%B-N@stanene (pattern-a) using HSE06 level of theory. Total energy verses time step of the AIMD Simulations carried out in different doped systems at room temperature. Phonon dispersion of different doped stanene. TDOS/PDOS under strains for some of these structures are presented. Electrostatic potentials of the system, structural parameters and band gap values are tabulated.

**Table of Content (TOC)**

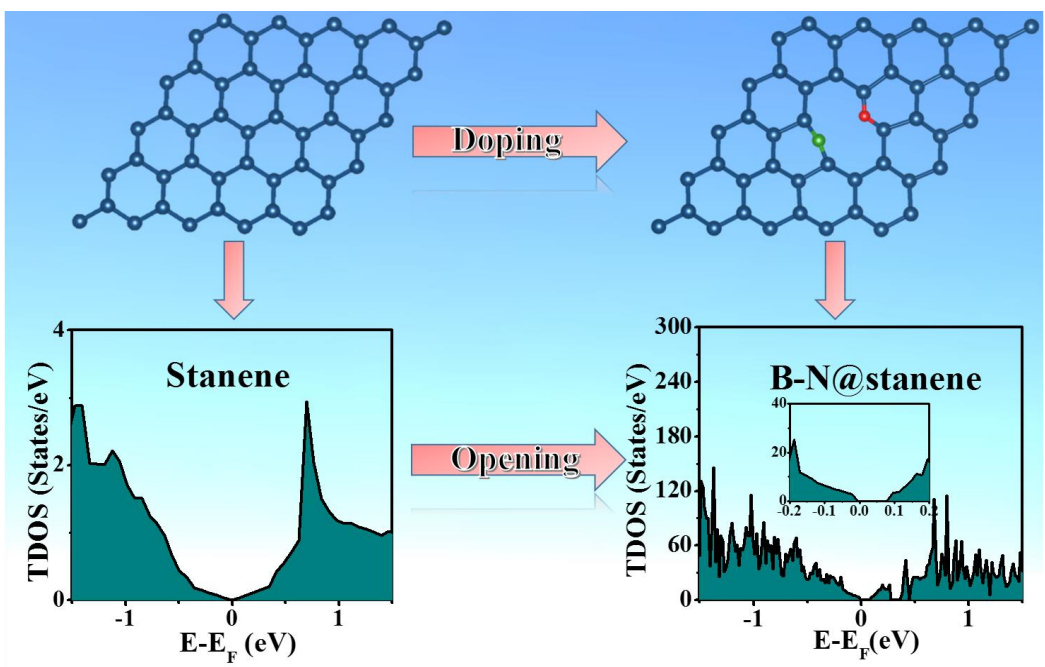